\documentclass[a4paper,12pt]{iopart}

\usepackage{graphicx}
\usepackage{amssymb}
\usepackage{iopams}
\usepackage{cite}
\usepackage{subfigure}

\newcommand{\beq}{\begin{equation}}
\newcommand{\eeq}{\end{equation}}
\newcommand{\baq}{\begin{eqnarray}}
\newcommand{\eaq}{\end{eqnarray}}

\def\remove#1{{[Remove: {\it #1}]}}
\def\remove#1{}
\def\refsub{{\rm ref}}

\begin{document}

\begin{flushright}
\begin{footnotesize}Imperial/TP/09/AC/01\end{footnotesize}
\end{flushright}

\title{Non-Gaussianity from resonant curvaton decay}

\author{Alex Chambers$^a$, Sami Nurmi$^b$ and Arttu Rajantie$^a$}
\address{$^a$ Department of Physics, Imperial College London, London SW7 2AZ, United Kingdom\\
$^b$Institute for Theoretical Physics, University of Heidelberg, 69120 Heidelberg, Germany
}
\ead{alex.chambers05@imperial.ac.uk, s.nurmi@thphys.uni-heidelberg.de and a.rajantie@imperial.ac.uk}

\bibliographystyle{JHEP}

\begin{abstract}
We calculate curvature perturbations in the scenario in which the
curvaton field decays into another scalar field via
parametric resonance. As a result of a nonlinear stage at the end of the resonance, standard
perturbative calculation techniques fail in this case.
Instead, we use lattice field theory simulations and the separate universe
approximation to calculate the curvature perturbation as a nonlinear function of
the curvaton field.
For the parameters tested, the generated perturbations are highly non-Gaussian
and
not well approximated by the usual $f_{\rm NL}$ parameterisation.
Resonant decay plays an important role in the curvaton scenario and can have a substantial effect on the resulting perturbations.
\end{abstract}

\section{Introduction}
Observations of the cosmic microwave background radiation
are consistent with Gaussian perturbations~\cite{Komatsu:2008hk}, but there
are
tantalising hints of non-Gaussianity~\cite{Yadav:2007yy} at a level that
would be clearly observable with the Planck satellite and other future
experiments~\cite{Komatsu:2009kd}.
This would rule out the simplest inflationary models with slowly rolling scalar
fields, which can only produce very-nearly-Gaussian perturbations
\cite{Maldacena:2002vr,Bartolo:2001cw,Acquaviva:2002ud,Seery:2005gb}.

Models which can generate large non-Gaussianities either during, or
at the end of, inflation include those with non-canonical
Lagrangians
\cite{Chen:2006nt,ArkaniHamed:2003uz,Alishahiha:2004eh,Seery:2005wm,Creminelli:2003iq,Langlois:2008qf,Arroja:2008yy,Chen:2009bc,Arroja:2009pd},
those where there is breakdown of slow-roll dynamics during
inflation due to sharp features in the potential
\cite{Chen:2006xjb,Chen:2008wn}, those with multiple fields with
specific inflationary trajectories in the field space
\cite{Bernardeau:2002jy,Bernardeau:2002jf,Lyth:2005qk,Salem:2005nd,Sasaki:2008uc,Naruko:2008sq,Byrnes:2008wi,Byrnes:2008zy,Byrnes:2009qy}
and those where additional light scalars affect the dynamics of
perturbative or non-perturbative inflaton decay \cite{
Kofman:2003nx,Dvali:2003em,Ichikawa:2008ne,Byrnes:2008zz,Enqvist:2004ey,Enqvist:2005qu,Barnaby:2006cq,Barnaby:2006km,Kohri:2009ac,Chambers:2007se,Chambers:2008gu,Bond:2009xx}.

A well-known example of a multi-field model where large
non-Gaussianities can be generated after the end of inflation is the
curvaton scenario
\cite{PhysRevD.42.313,Linde:1996gt,Enqvist:2001zp,Lyth:2001nq,Moroi:2001ct}.
In this model, the primordial perturbations arise from the curvaton field, a
scalar field which is light relative to the Hubble rate during inflation but,
unlike the inflaton,
gives a subdominant contribution to the energy density.

After inflation the relative
curvaton contribution to the total energy
density increases. This affects the expansion of space,
and the curvaton perturbations become imprinted on the metric fluctuations.
The standard adiabatic hot big bang era is recovered when the
curvaton eventually decays and thermalises with the existing
radiation. The mechanism can be seen as a conversion of initial
isocurvature perturbations into adiabatic curvature perturbations
during the post-inflationary epoch. If the curvaton remains
subdominant even at the time of its decay, its perturbations need to
be relatively large to yield the observed amplitude of primordial
perturbations. Consequently, in this limit the curvaton scenario
typically generates significant non-Gaussianities
\cite{Lyth:2002my,Bartolo:2003jx,Gordon:2003hw,Enqvist:2005pg,Malik:2006pm,Sasaki:2006kq,Valiviita:2006mz,Assadullahi:2007uw}.

Almost all the analysis of the curvaton scenario is based on the
assumption of a perturbative curvaton decay. As pointed out in refs.~\cite{BasteroGil:2003tj,Enqvist:2008be}, it is,
however, possible that the curvaton
decays through a non-perturbative process analogous to inflationary
preheating \cite{Traschen:1990sw,Kofman:1994rk,Kofman:1997yn}.
This is a natural
outcome in models where the curvaton is coupled to other scalar
fields which acquire effective masses proportional to the value of the curvaton field.

During this short period of non-equilibrium physics, part of the
curvaton condensate decays rapidly into quanta of other light
scalar fields. In analogy to inflationary preheating, the decay is
not complete and the remaining curvaton particles have to decay
perturbatively. The properties of primordial perturbations generated in the
curvaton scenario depend sensitively on the dynamics after the end
of inflation until the time of the curvaton decay
\cite{Dimopoulos:2003ss,Enqvist:2009zf,Lemoine:2009is}. As the
preheating dynamics are highly nonlinear, it is natural to ask if
this stage could generate large non-Gaussianities.

Our aim in this
work is to calculate the amplitude and non-Gaussianity of
perturbations from curvaton preheating. To do this we use the
separate universe approximation
\cite{Starobinsky:1986fxa,Salopek:1990jq,Sasaki:1995aw,Lyth:2004gb}
and classical lattice field theory simulations
\cite{Khlebnikov:1996mc,Prokopec:1996rr}. This method was applied to
inflationary preheating in
refs.~\cite{Chambers:2007se,Chambers:2008gu,Bond:2009xx}. We solve
coupled field and Friedmann equations to determine the expansion of the
universe during the non-equilibrium dynamics and the fraction of the
curvaton energy which is converted into radiation. From these, we
calculate the curvature perturbation as a function of the local
value of the curvaton field.
We find that, at least for our choice of
parameters, this dependence is highly nonlinear,
implying very high levels of non-Gaussianity.

The structure of the paper is as follows. In section
\ref{sect:evolveofperts} we review the curvaton model and derive
useful general results. In section \ref{sect:analytic} we present an
analytic calculation of production of curvature perturbations using
linearised approximation of preheating dynamics, and demonstrate
that the answers depend on quantities that are not calculable in
linear theory. A full nonlinear computation is, therefore, needed.
In section \ref{sect:sims} we show how this can be done, and compute
the amplitude and non-Gaussianity of the perturbations for three sets
of parameters. Finally, we present our conclusions in section
\ref{sect:conclude}.

\section{General theory}
\label{sect:evolveofperts}

\subsection{Model}
\label{sect:model}
We study the curvature perturbations generated in a simple curvaton model with
canonical kinetic terms and the potential
 \beq
 \label{V}
 V(\phi,\sigma,\chi)=V(\phi)+\frac{1}{2}m^2\sigma^2+\frac{1}{2}g^2\sigma^2\chi^2\
,
 \eeq
where $\phi$ is the inflaton, $\sigma$ is the curvaton and $\chi$ is
a scalar field which the curvaton will decay into. During
inflation the inflaton potential $V(\phi)$ dominates the energy
density. We assume standard slow roll inflation, so the inflaton
field rolls down its potential until it reaches a critical value
$\phi_*$ at which the slow roll conditions fail and inflation ends.
As usual, the perturbations of the inflaton field generate curvature
perturbations, but we assume that their amplitude is negligible.
This requires $H/\sqrt{\epsilon}\ll 10^{-5}$, where $H$ is the
Hubble rate and $\epsilon$ is the slow roll parameter, both of which
are determined by the inflaton potential $V(\phi)$.

Instead, we assume that the observed primordial perturbations are
generated solely by the curvaton field $\sigma$, which should be
light during inflation so that it develops nearly scale-invariant
quantum fluctuations similar to those of the inflaton field.
The curvaton field remains nearly frozen at the value $\sigma_*$ it
had at the end of inflation, until the time $t_{\rm osc}$ at which
the Hubble parameter has decreased to $H\sim m$ and the curvaton
starts to oscillate around the minimum of its potential.
These fluctuations contribute to the energy density, and, therefore, they affect the
expansion of the universe and generate curvature perturbations.

Eventually, the curvaton $\sigma$ decays into lighter degrees of
freedom. In the
standard curvaton scenario this decay is assumed to take place
perturbatively. The interaction term in (\ref{V}) only allows pair
annihilations, the rate of which falls quickly when the universe
expands. Some of the curvaton particles would, therefore, survive
until today, behaving as dark matter. While this might be an
interesting scenario to study, we follow most of the literature and
assume that the curvaton is also coupled to other fields. In
particular, a Yukawa coupling to a light (possibly Standard Model)
fermion field $\psi$ of the form
\begin{equation}
\label{equ:Yukawa}
{\cal L}_{\rm Yukawa}=h\sigma\bar\psi\psi
\end{equation}
would give the decay rate~\cite{Kofman:1997yn}
\begin{equation}
\label{equ:Yukawarate}
\Gamma=\frac{h^2m}{8\pi}\ .
\end{equation}
Without further knowledge of the curvaton couplings, the decay rate
$\Gamma$ appears as a free phenomenological parameter, which is
bounded from below because
the curvaton cannot decay arbitrarily late
without spoiling the success of the hot big bang cosmology. The most
conservative lower bound for the decay time is given by
nucleosynthesis: $\Gamma\gtrsim T^2_{\rm BBN}/M_{\rm Pl}$
\cite{Lyth:2003dt}.

In this work, we assume that the other scalar field $\chi$ in
(\ref{V}) is heavy during inflation, $g\sigma \gtrsim H$, so that
its value remains close to zero and it develops no long-range
perturbations. It was recently shown in ref.~\cite{Enqvist:2008be}
that in this case some of the curvaton particles can decay via a
parametric resonance into quanta of the $\chi$ field. The resonance
does not destroy all the curvaton particles, so a Yukawa coupling is
still needed to allow those which remain to decay perturbatively.
However, it reduces their density significantly.

\subsection{Curvature perturbations}
\label{sect:evolveofperts_preheating}

Using the $\delta N$
approach, the curvature perturbation $\zeta$ on superhorizon scales
is given by the expression \cite{Salopek:1990jq}
 \begin{equation}
 \label{equ:deltan}
 \zeta=\delta \ln a\Big|_{\rho},
 \end{equation}
where $a|_{\rho}$ is the scale factor at some fixed energy density
$\rho$, and $\delta$ denotes the difference from the mean value over
the whole currently observable universe. The scale factor $a$ is
normalised to be constant at some earlier flat hypersurface, which
we choose to be at the end of inflation. This measures the
difference in the integrated expansion between Friedmann--Robertson--Walker (FRW) solutions with different
initial conditions which are evolved until the same final energy density
$\rho$. In our case, the variations of initial conditions arise from
the super-horizon fluctuations of the curvaton field
$\delta\sigma_{*}$ produced during inflation. This means that the
curvature perturbation is a local function of the curvaton field
fluctuations $\zeta=\zeta(\delta\sigma_*)$. As we know the statistics of
$\delta\sigma_*$, this determines the statistics of the curvature perturbation $\zeta$ completely.
In this paper we
compute this function.

Observations show that the primordial perturbations were nearly
Gaussian, which implies that $\zeta(\delta\sigma_*)$ should be
approximately linear. Therefore, it is natural to Taylor expand
(\ref{equ:deltan}) as
   \baq
  \label{zeta_exp}
  \zeta&=&({\rm ln}\,a)'\Big|_{\rho}\delta\sigma_{*}+\frac{1}{2}({\rm ln}\,a)''\Big|_{\rho}\delta\sigma_{*}^2
  +\ldots\ ,
  \eaq
where the prime denotes a partial derivative with respect to the
curvaton value at the end of inflation $\sigma_{*}$ evaluated at
constant final energy density $\rho$. This implies that, to leading
order in $\delta\sigma_{*}$, the power spectrum of the curvature
perturbations is
\begin{equation}
 \label{eqn:powerzeta}
 \mathcal{P}_\zeta (k)=\left[(\ln a)'\Big|_{\rho}\right]^2\mathcal{P}_\sigma(k),
\end{equation}
where $\mathcal{P}_\sigma(k)$ is the power spectrum of the curvaton
field. Assuming the curvaton perturbations at the end of inflation
$\delta\sigma_{*}$ are Gaussian, the expansion (\ref{zeta_exp}) is
of the form \cite{Komatsu:2000vy}
  \beq
  \label{fnl_local}
  \zeta=\zeta_{\rm g}+\frac{3}{5}f_{\rm NL}\zeta_{\rm g}^2+\ldots\ ,
  \eeq
where $\zeta_{\rm g}$ is a Gaussian field and $f_{\rm NL}$ is a
position independent constant. This defines the local nonlinearity
parameter $f_{\rm NL}$ and, according to (\ref{zeta_exp}), the $\delta
N$ formalism gives \cite{Lyth:2005fi}
  \beq
  \label{fnl}
  f_{\rm NL}=\frac{5}{6}\frac{({\rm ln}\,a)''}{({\rm
ln}\,a)'{}^2}\left|\rule{0 pt}{3 ex}\right._{\rho}\ .
  \eeq
Neglecting the inherent non-Gaussianity of the curvaton
perturbations $\delta\sigma_{*}$ yields an error proportional to
slow-roll parameters. This is irrelevant in the limit $|f_{\rm
NL}|\gg 1$, which is the main focus of this work. The WMAP data and
other recent observations give the constraint $|f_{\rm NL}|\lesssim
100$~\cite{Komatsu:2008hk,Yadav:2007yy,Komatsu:2009kd,Senatore:2009gt}.

In this paper we focus on the case in which the curvaton field decays
into $\chi$ particles through a parametric resonance, at the
end of which the fields undergo a period of potentially very complicated, nonlinear, non-equilibrium dynamics. Afterwards, the fields equilibrate,
and we assume that eventually the universe behaves as a
mixture of non-interacting matter and radiation. This assumption is
checked using simulations in section \ref{sect:sims} and we find it to be accurate enough for our current purposes. Therefore, we
parameterise the total energy density as a sum of matter and radiation components
\begin{equation}
  \label{rho_split}
  \rho=\rho_\refsub \left[r_\refsub \left({a_\refsub \over a}\right)^3+(1-r_\refsub )\left({a_\refsub \over
  a}\right)^4\right]\ ,
\end{equation}
where $a_\refsub $, $\rho_\refsub $ and $r_\refsub =\rho_{m,\rm
ref}/\rho_\refsub $ are the scale factor, energy density and matter
fraction at some arbitrary reference time well after the resonance
is over.\footnote{ We define $r=\rho_m/\rho$ following
ref.~\cite{Lyth:2001nq}. This differs from the variable $r$ used in
ref.~\cite{Lyth:2002my} by a factor of $3/4$ in the limit $r\ll 1$.}

We assume that the inflaton field has either decayed into ultrarelativistic degrees of freedom or
is itself ultrarelativistic,
so that it contributes to the radiation component. The $\chi$ field is also ultrarelativistic;
the only degree of
freedom that contributes to the matter component is the curvaton.

If the resonance destroyed all of the curvaton particles, $r_\refsub$ would be zero, but, in practice, some
curvatons are left over.
We assume $r_\refsub \ll 1$ which corresponds to
the curvaton being subdominant at the end of the resonance. We can then
rearrange (\ref{rho_split}) to give, at leading order in
$r_\refsub $,
the scale factor at energy density $\rho$
\begin{equation}
 \label{eqn:rhosplit_a}
 \ln a = \ln a_\refsub  + {1\over 4}\left[\ln{\rho_\refsub \over \rho}+r_\refsub \left(\left({\rho_\refsub \over\rho}\right)^{1/4}-1\right)\right]\ ,
\end{equation}
where we have assumed that the curvaton is still subdominant at this time. That is
 \begin{equation}
 \label{eqn:revolve}
  r\equiv r_\refsub \left({\rho_\refsub \over\rho}\right)^{1/4}\ll 1.
\end{equation}

The curvature perturbation is given by combining
(\ref{equ:deltan}) with (\ref{eqn:rhosplit_a}). In general,
$a_\refsub $, $\rho_\refsub $ and $r_\refsub $ all vary between one
separate universe and another. In our case, the variation is entirely
due to fluctuations of $\sigma_*$, the value of the curvaton field
during inflation. In order to calculate the curvature perturbation
using (\ref{zeta_exp}) we differentiate (\ref{eqn:rhosplit_a}) with
respect to $\sigma_{*}$ keeping the final energy density $\rho$
fixed. This gives
\begin{eqnarray}
 \label{eqn:lnap}
\fl
 (\ln a)'\Big|_{\rho}&=&(\ln a_\refsub )'+{1 \over 4}
 \left[ \frac{\rho'_\refsub }{\rho_\refsub }+
 r_\refsub \frac{\rho'_\refsub }{4 \rho^{1/4}\rho_\refsub ^{3/4}}+
 r_\refsub '\left(\left({\rho_\refsub \over\rho}\right)^{1/4}-1\right)
 \right]\nonumber\\
\fl
 &=&(\ln a_\refsub )'+{1 \over 4}
 \left[ \left(1+\frac{r}{4}\right)\frac{\rho'_\refsub }{\rho_\refsub }+
 (r-r_\refsub ){r_\refsub '\over r_\refsub }
 \right],\\
\fl
 \label{eqn:lnapp}
 (\ln a)''\Big|_{\rho}&=&(\ln a_\refsub )''+{1 \over 4}
 \left[ \frac{\rho''_\refsub }{\rho_\refsub }-\left(\frac{\rho'_\refsub }{\rho_\refsub }\right)^2+
 \frac{r_\refsub }{4}\left(\frac{\rho_\refsub }{\rho}\right)^{1/4}
\left( \frac{\rho''_\refsub }{\rho_\refsub }-\frac{3}{4}\left(\frac{\rho'_\refsub }{\rho_\refsub }\right)^2\right)\right.  \nonumber
\\
\fl
  &&
 \left. +r_\refsub '\frac{\rho'_\refsub }{2\rho^{1/4}\rho_\refsub ^{3/4}}
 +r''_\refsub \left( \left(\frac{\rho_\refsub }{\rho}\right)^{1/4}-1\right)
 \right]\nonumber\\
\fl
 &=&(\ln a_\refsub )''+{1 \over 4}
 \left[\rule{0pt}{3ex}\right. \frac{\rho''_\refsub }{\rho_\refsub }-\left(\frac{\rho'_\refsub }{\rho_\refsub }\right)^2+
 \frac{r}{4}\left( \frac{\rho''_\refsub }{\rho_\refsub }-\frac{3}{4}\left(\frac{\rho'_\refsub }{\rho_\refsub }\right)^2\right)\nonumber\\
\fl
 &&+
 \frac{r}{2} \frac{r'_\refsub}{r_\refsub }\frac{\rho_\refsub '}{\rho_\refsub
 }+
 (r-r_\refsub ){r_\refsub ''\over r_\refsub }
 \left]\rule{0pt}{3ex}\right. .
\end{eqnarray}

The remaining curvatons decay within a fixed time $1/\Gamma$ where
the decay rate $\Gamma$ depends on the curvaton interactions. We
assume that all matter in the universe is
ultrarelativistic after the curvaton decay, so that the universe
becomes radiation dominated and evolves adiabatically. Following the
sudden decay approximation~\cite{Lyth:2001nq}, we assume that this
decay takes place instantaneously at a fixed energy density
$\rho_{\rm decay}$, which is determined by the curvaton decay rate
$\Gamma$. The final curvature perturbation is therefore given by
setting $\rho=\rho_{\rm decay}$ in (\ref{eqn:lnap}) and
(\ref{eqn:lnapp}), or equivalently $r=r_{\rm decay}$, where
\begin{equation}
 r_{\rm decay}=r_\refsub \left({\rho_\refsub \over\rho_{\rm decay}}\right)^{1/4}.
\end{equation}
The decay rate $\Gamma$  is
unknown, so we can treat $r_{\rm decay}$ as a free parameter.

\subsection{Standard perturbative decay}
\label{sect:evolveofperts_standard}

As a simple example of the calculation of perturbations, and to aid
comparison with our results for resonant curvaton decay, we
first summarise the standard perturbative calculation of perturbations generated in the curvaton
model~\cite{Lyth:2001nq,Lyth:2002my}.

This calculation ignores the resonance, so (\ref{rho_split}) is assumed to be valid
from the start of the curvaton oscillations, which we can therefore choose as our reference
time, $t_{\rm ref}=t_{\rm osc}$.
This time is determined by the condition $H\approx m$, which
implies that $\rho_{\rm ref}=\rho_{\rm osc}\approx m^2M_{\rm Pl}^2$
independently of $\sigma_*$. Furthermore, the scale factor
$a_\refsub =a_{\rm osc}$ is also independent of
$\sigma_*$ to good approximation, because the curvaton
contribution to the energy density is negligible during inflation. Perturbations are
therefore generated solely by derivatives of $r_\refsub =r_{\rm
osc}$, the matter fraction at the start of the oscillations, and (\ref{eqn:lnap}) and (\ref{eqn:lnapp}) simplify to
\begin{eqnarray}
 \label{eqn:lnap_pert}
 (\ln a)'\Big|_{\rho}&=&{r_{\rm decay} \over 4}{r_{\rm osc}'\over r_{\rm osc}}\ ,\\
 \label{eqn:lnapp_pert}
 (\ln a)''\Big|_{\rho}&=&{r_{\rm decay} \over 4}{r_{\rm osc}''\over r_{\rm osc}}\
 ,
\end{eqnarray}
where we have also assumed $r_{\rm osc}\ll r_{\rm decay}\ll 1$. Using the
results (\ref{rho_sigma_0}) and (\ref{sigma_0}) below, the
derivatives of the matter fraction $r_{\rm osc}=m^2\sigma^2_{\rm
osc}/2\rho_{\rm osc}$ read
\begin{equation}
\label{equ:rosc}
\frac{r_{\rm osc}'}{r_{\rm osc}}=\frac{2}{\sigma_{*}}\ ,\
\frac{r_{\rm osc}''}{r_{\rm osc}} =\frac{2}{\sigma_{*}^2}
\end{equation}
giving the simple result~\cite{Lyth:2002my}
\begin{equation}
 \label{eqn:fnl_pert}
 f_{\rm NL}=\frac{5}{3}\frac{1}{r_{\rm decay}}\ .
\end{equation}

\section{Linearised calculations}
\label{sect:analytic}
\subsection{Resonance}
Our general result in (\ref{eqn:lnap}) and (\ref{eqn:lnapp})
depends on the quantities $a_{\rm ref}$, $\rho_{\rm ref}$ and
$r_{\rm ref}$ and their derivatives. Ideally, we would like to be
able to calculate them analytically for general parameter values. We
first attempt to carry out this calculation in linear theory of
resonance \cite{Kofman:1997yn}, i.e. neglecting the backreaction of
particles produced during the resonance. This approximation gives a
clear physical picture of the resonance, and also an accurate
quantitative description of many aspects of it. However, we find
that it is not suitable for calculating the curvature perturbation,
and, therefore, this calculation acts, ultimately, as a motivation for
the nonlinear approach in section~\ref{sect:sims}.

As we assume the $\chi$ field is heavy during inflation and has a
vanishing vacuum expectation value, and that the universe becomes
radiation dominated after the end of inflation, the equation of
motion for the curvaton is given by
 \beq
 \label{sigma_eom}
 \ddot{\sigma}+\frac{3}{2t}\,\dot{\sigma}+m^2\sigma=0 \ ,
 \eeq
which can easily be put into the canonical form of a Bessel
equation. The general solution of (\ref{sigma_eom}) which remains
bounded as $t\rightarrow 0$ is given by
  \beq
  \label{sigma(t)}
  \sigma(t)=2^{1/4}\Gamma(5/4)\sigma_{*}\,\frac{J_{1/4}(mt)}{(mt)^{1/4}}\ ,
  \eeq
where $J_{1/4}(x)$ is a Bessel function of the first kind and
$\sigma(0)=\sigma_{*}$ denotes the initial curvaton value set by the
dynamics during inflation. In the model (\ref{V}) that we
consider here, the value of the curvaton field is constrained by
  \beq
  \label{sigma_constraint}
  \frac{H_{*}^2}{g^2}< \sigma_{*}^2 < \frac{H_{*}^2M_{\rm Pl}^2}{m^2}\ ,
  \eeq
where the upper bound comes from the subdominance of the curvaton
and the lower bound is required to keep the $\chi$ field massive
during inflation and to enable broad parametric resonance
\cite{Enqvist:2008be}.

The curvaton remains nearly frozen at the value $\sigma_{*}$ until
the Hubble parameter decreases to $H\sim m$ and the field starts
oscillate around the minimum of its potential. We assume the
curvaton is still subdominant at this time and obeys the
solution (\ref{sigma(t)}). After the onset of oscillations,
$mt\gtrsim 1$, (\ref{sigma(t)}) can be approximated by the
asymptotic expression
  \beq
  \label{sigma_osc}
  \sigma(t) \approx  \frac{2^{3/4}\Gamma(5/4)}{\pi^{1/2}}\frac{\sigma_{*}}{(mt)^{3/4}}\,{\rm
  sin}\left(mt+\frac{\pi}{8}\right)\ .
  \eeq
Following the notation of \cite{Kofman:1997yn}, we choose to
normalise the scale factor to unity at $t_{\rm osc}=3\pi/(8 m)$
  \beq
  \label{a}
  a=\left(\frac{t}{t_{\rm osc}}\right)^{1/2}=\left(\frac{8 m t}{3 \pi}\right)^{1/2}\ .
  \eeq
Up to corrections ${\cal{O}}(H/m)$, the energy density of the
oscillating curvaton is given by
  \beq
  \label{rho_sigma_0}
  \rho_{\sigma}=\frac{1}{2}\frac{m^2\sigma_{\rm osc}^2}{a^3}\equiv
  \frac{\rho_{\sigma,{\rm osc}}}{a^{3}}\ ,
  \eeq
where $\sigma_{\rm osc}$ is the envelope of the oscillatory solution
(\ref{sigma_osc}) at $t_{\rm osc}$
  \beq
  \label{sigma_0}
  \sigma_{\rm osc}=\frac{8\,\Gamma(5/4)}{3^{3/4}\pi^{5/4}}\,\sigma_{*}\approx 0.76\,\sigma_{*}\ .
  \eeq
The time $t_{\rm osc}$ appears as an unphysical reference point in
our analysis but as it formally corresponds to a quarter of the
first oscillation cycle it can also be thought of as the beginning
of curvaton oscillations, hence the subscript.

The coherent curvaton oscillations induce a time-varying effective
mass for the $\chi$ field which can lead to copious production of
$\chi$ particles as discussed in \cite{Enqvist:2008be}. The process
is analogous to the standard inflationary preheating scenario
\cite{Kofman:1994rk,Kofman:1997yn,Greene:1997fu}, except that the
universe is dominated by radiation instead of non-relativistic
matter during the resonance. The resonant curvaton decay is
efficient if
  \begin{equation}
  \label{q}
  q=\frac{g^2 \sigma_{\rm osc}^2}{4m^2}\gg 1\ ,
  \end{equation}
which corresponds to broad resonance bands in momentum space. For
curvaton values in the range (\ref{sigma_constraint}) this condition
is always satisfied, as one can immediately see by taking into
account the lightness of the curvaton during inflation, $m\ll H_{*}$.

Using the solution (\ref{sigma_osc}) and applying the analytic
methods developed to analyse particle production during the
parametric resonance \cite{Kofman:1997yn}, the comoving number
density of $\chi$ particles at late times ($t\gg
t_{\rm osc}$) can be estimated as
  \beq
  \label{n(t)}
  n_{\chi}(t)\approx \frac{k_{*}^3}{64 \pi^2 a^3 \sqrt{\mu m t}} e^{2m \mu t}\ ,
  \eeq
where $k_{*}=(g m \sigma_{\rm osc})^{1/2}$ and $\mu\sim 0.14$ is an effective growth index. Due to the stochastic nature of parametric resonance
in expanding space, the actual growth index depends very
sensitively on other parameters and varies rapidly between $0$ and $0.28$ between
one cycle of oscillation and another.

As the number density of $\chi$ particles grows exponentially, the
contribution to the effective curvaton mass
$g^2\langle\chi^2\rangle$ eventually comes to dominate over the bare
mass $m^2$. The time $t_{\rm br}$ when the two contributions are equal---after which the
backreaction of the $\chi$ particles can no longer be neglected---is
found by equating the number density $n_{\chi}\sim
\langle\chi^2\rangle g\sigma$ with $m^2\sigma/g$. Using the result
(\ref{n(t)}), this yields an equation for $t_{\rm br}$
  \beq
  t_{\rm br}\approx \frac{1}{4m\mu}{\rm ln}\left(\frac{10^5\mu m(mt_{\rm
br})^{5/2}}{g^5\sigma_{\rm osc}}\right)\ ,
  \eeq
whose solution is given by the $-1$ branch of Lambert W-function
  \beq
  \label{t_br}
  t_{\rm
br}\approx-\frac{5}{8m\mu}W_{-1}\left(-10^{-2}g^{8/5}\mu^{3/5}q_{\rm
osc}^{1/5}\right)\ .
  \eeq
Here $q_{\rm osc}$ denotes the value of the resonance parameter
(\ref{q}) at the beginning of oscillations. The result is consistent
with the assumption $t\gg t_{\rm osc}$ for all $\mu$ provided that
we place the mild constraint $g^{8/5}q_{\rm osc}^{1/5}\lesssim 10$,
which corresponds to $t_{\rm br}\gtrsim 10\,t_{\rm osc}$. In this
limit (\ref{t_br}) can be approximated to reasonable
accuracy by
  \beq
  \label{t_br_appr}
  t_{\rm br}\sim \frac{1}{8m\mu}{\rm ln}\left(10^{10}g^{-8}\mu^{-3}
q_{\rm osc}^{-1}\right)+{\cal {O}}\left(m^{-1}\,{\rm ln}(m t_{\rm
br})\right)\ .
  \eeq

The value of the scale factor at the time of backreaction is found
by substituting (\ref{t_br}) into (\ref{a})
  \begin{equation}
  \label{a_br}
  a_{\rm br}\approx\left(-\frac{5}{3\pi\mu}\right)^{1/2}
  W_{-1}^{1/2}\left(-10^{-2}g^{8/5}\mu^{3/5}q_{\rm osc}^{1/5}\right)\ .
  \end{equation}
The derivatives of ${\rm
ln}\,a_{\rm br}$ with respect to $\sigma_{*}$, which are needed for
computing the curvature perturbation, are
  \baq
  \label{lnabr'}
  ({\rm ln}\,a_{\rm br})'&\approx &-\frac{1}{2}({\rm ln}\,\mu)'-\frac{1}{3\pi\mu a_{\rm br}^2\sigma_{*}}
  \left(1+{\cal {O}}(a_{\rm br}^{-2})\right)
  \\
  \label{lnabr''}
  ({\rm ln}\,a_{\rm br})''&\approx&-\frac{1}{2}({\rm ln}\,\mu)''+
  \frac{1}{3\pi\mu a_{\rm br}^2\sigma_{*}^2}\left(1+{\cal {O}}(a_{\rm
br}^{-2})\right)\ ,
  \eaq
where $'\equiv\partial/\partial{\sigma}_{*}$ and the relation
$\partial^n/\partial{\sigma}_{*}^n=(\sigma_{\rm
osc}/\sigma_{*})^n\,\partial^n/\partial\sigma_{\rm osc}^n$ following
from (\ref{sigma_0}) has been used. The terms involving
derivatives of the effective growth index $\mu$ would be very
difficult to estimate analytically and we therefore leave them in an
implicit form.

Approximate energy conservation $\frac{1}{2}\,m^2\sigma_{\rm
br}^2+m^2\sigma_{\rm br}^2=1/2\, m^2\sigma_{\rm osc}^2 a_{\rm br}^{-3}$
gives an estimate
  \beq
  \label{q_br}
  q_{\rm br}\approx \frac{1}{12}\frac{g^2\sigma_{\rm osc}^2}{m^2a_{\rm br}^3} = \frac{1}{3}\frac{q_{\rm osc}}{a_{\rm br}^3}
  \eeq
for the resonance parameter (\ref{q}) at the onset of backreaction.
If $q_{\rm br}\gtrsim 1$, the resonance does not terminate
immediately at $t_{\rm br}$. Instead, the effective curvaton mass $m^2_{\rm eff}=
g^2\langle\chi^2\rangle$ starts to grow exponentially, which
soon brings the resonance to end. The nonlinear stage of the
resonance after $t_{\rm br}$ therefore makes only a small
contribution to the total duration of the resonance
\cite{Kofman:1997yn} which can be reasonably well estimated by the
duration of the linear stage $t_{\rm br}$ given by (\ref{t_br}).

\subsection{Curvature perturbation}

It is not possible to describe the backreaction and the
equilibration of the fields using linear theory.
To carry out the linearised calculation, we therefore simply assume that
the resonance ends instantaneously at time $t_{\rm br}$, and that
some fraction $\xi$ of the curvaton energy is transferred into
radiation. The matter and radiation energy densities immediately
after backreaction are therefore given by
  \baq
  \label{rho_m1}
  \rho_{m,{\rm br}}&=&(1-\xi)\frac{\rho_{\sigma,\rm osc}}{a_{\rm br}^3}\ ,\\
  \label{rho_r1}
  \rho_{r,{\rm br}}&=&\frac{\rho_{r,\rm osc}}{a_{\rm br}^4}+\xi\frac{\rho_{\sigma,\rm osc}}{a_{\rm
  br}^3}\ .
  \eaq

In order to calculate the curvature perturbation, we choose this backreaction time as our reference time in (\ref{rho_split}). Therefore we have
$a_{\rm ref}=a_{\rm br}$, and
\begin{eqnarray}
\label{eqn:rhobr}
\rho_{\rm ref}&=&\rho_{\rm br}=\rho_{m,{\rm br}}+\rho_{r,{\rm br}}
=\frac{\rho_{r,\rm osc}}{a_{\rm br}^4}+\frac{\rho_{\sigma,\rm osc}}{a_{\rm
  br}^3}
  \approx \frac{\rho_{\rm osc}}{a^4_{\rm br}}\left(1+r_{\rm osc}(a_{\rm br}-1)\right)\ ,
\\
\label{eqn:rbr}
 r_{\rm ref}&=&r_{\rm br}=\frac{\rho_{m,{\rm br}}}{\rho_{m,{\rm br}}+\rho_{r,{\rm br}}}
 =
 \frac{(1-\xi)\rho_{\sigma,\rm osc}a_{\rm br}}{\rho_{r,\rm osc}+\rho_{\sigma,\rm osc}a_{\rm br}}
 \approx (1-\xi)r_{\rm osc}a_{\rm br}\ ,
\end{eqnarray}
where $r_{\rm osc}=\rho_{\sigma,{\rm osc}}/\rho_{\rm osc}\approx
\rho_{\sigma,{\rm osc}}/\rho_{r,\rm osc}$
 and we have assumed that $r_{\rm
osc}a_{\rm br}\ll 1$, which is equivalent to having $r\ll 1$
throughout the resonance. Substituting these into (\ref{eqn:lnap}) and (\ref{eqn:lnapp}), we can work out the
expression for the curvature perturbation. Using the leading terms
in (\ref{lnabr'}), (\ref{lnabr''}), we find
  \baq
  \label{lna'}
\fl
  ({\rm ln}\,a)'\Big|_{\rho}&=&\frac{r}{4}\left[\rule{0pt}{4ex}\right.
  \frac{\xi}{1-\xi}\frac{a_{\rm
br}}{a}\left(\frac{2}{\sigma_{{*}}}-\frac{1}{2}\frac{\mu'}{\mu}
  +\frac{\xi'}{\xi}\right)+\frac{2}{\sigma_{*}}-\frac{\xi'}{1-\xi}
  \left]\rule{0pt}{4ex}\right.\ ,
  \\
  \label{lna''}
\fl
\nonumber
  ({\rm ln}\, a)''\Big|_{\rho}&=&\frac{r}{4}\left[\rule{0pt}{4ex}\right.
  \frac{\xi}{1-\xi}\frac{a_{\rm br}}{a}\left(\rule{0pt}{3ex}\right.
  \frac{2}{\sigma_{*}^2}+\frac{4}{\sigma_{*}}\left(\frac{\xi'}{\xi}-\frac{1}{2}\frac{\mu'}{\mu}\right)+
  \frac{\xi''}{\xi}+
  \frac{(\mu^{-1/2})''}{\mu^{-1/2}}
  -\frac{\xi'\mu'}{\xi\mu}
  \left)\rule{0pt}{3ex}\right.
  \\
\fl
  &&
  +\frac{2}{\sigma_{*}^2}-\frac{4\xi'}{\sigma_*(1-\xi)}-\frac{\xi''}{1-\xi}
  \left]\rule{0pt}{4ex}\right.\ .
  \eaq
By setting $\xi=0$ in the above expressions, the standard curvaton
results (\ref{eqn:fnl_pert}) for a perturbative decay are recovered.

The terms scaling as $a^{-1}$ in (\ref{lna'}) and (\ref{lna''})
represent the contribution due to radiation inhomogeneities created
by the curvaton decay. If the curvaton particles left over after the resonance have a long perturbative lifetime, this\remove{
The resonant curvaton decay is not complete
and the resonance must be eventually followed by a perturbative
decay which generates the Standard Model degrees of freedom that
make up the currently observable universe. This happens roughly at
$\Gamma\sim H$ where $\Gamma$ is the model dependent perturbative
decay width of the curvaton. If $\Gamma\ll H_{\rm br}$, the
perturbative decay takes place relatively long time after the the
end of resonance. In this limit the inhomogeneous radiation}
component effectively redshifts away and the expressions
(\ref{lna'}) and (\ref{lna''}) reduce to
  \baq
  \label{lna'_as}
  ({\rm ln}\,a)'\Big|_
  {\rho}&\sim&\frac{r}{2\sigma_{*}}\left(1-\frac{\sigma_{*}\xi'}{2(1-\xi)}
  \right)
  \\
  \label{lna''_as}
  ({\rm ln}\, a)''\Big|_{\rho}&\sim&\frac{r}{2\sigma_{*}^2}\left(
  1-\frac{2\sigma_{*}\xi'}{1-\xi}-\frac{\sigma_{*}^2\xi''}{2(1-\xi)}
  \right)\ .
  \eaq

The nonlinearity parameter $f_{\rm NL}$ can be computed using
(\ref{fnl}). In the late time limit, we find using (\ref{lna'_as})
and (\ref{lna''_as}) ,
  \beq
  \label{fnl_as}
  f_{\rm NL}\sim\frac{5}{3 r}\left(1-\frac{2\sigma_{*}\xi'}{1-\xi}-\frac{\sigma_{*}^2\xi''}{2(1-\xi)}\right)
  \left(1-\frac{\sigma_{*}\xi'}{2(1-\xi)}\right)^{-2}\  .
  \eeq

To make use of these results, one would have to know the first and second derivatives of $\mu$ and $\xi$, which, unfortunately, are difficult to calculate. In principle, the exponential growth rate $\mu$ is calculable in linearised theory, but obtaining its derivatives reliably is hard because it is an averaged quantity that describes a stochastic process. In contrast, $\xi$ is fully determined by the nonlinear dynamics at the end of the resonance, and, therefore, it is not possible to calculate it using linearised equations.

We conclude that the perturbations generated by resonant curvaton decay are not calculable using linear theory. Instead, we have to carry out a fully nonlinear calculation using lattice field theory methods.

\section{Simulations}
\label{sect:sims}
\subsection{Simulation method}

In section~\ref{sect:analytic} we saw that the curvature perturbations produced by the resonance cannot be calculated using linear theory. Therefore, we adopt a completely different approach.
We use nonlinear three-dimensional classical lattice field theory simulations\footnote{We use a modification of the corrected version of the code
used in refs. \cite{Chambers:2007se} and \cite{Chambers:2008gu} described in the
erratum of \cite{Chambers:2008gu}.} to determine how the scale
factor evolves as the fields fall out of equilibrium at the end of preheating and to
compute the required quantities $a_{\rm ref}$, $\rho_{\rm ref}$ and $r_{\rm ref}$ directly.
This is a standard method of solving the field dynamics in such systems
\cite{Khlebnikov:1996mc,Prokopec:1996rr,Felder:2000hq,Rajantie:2000fd,Rajantie:2000nj,Copeland:2002ku,Podolsky:2005bw,BasteroGil:2007mm},
and recently \cite{Chambers:2007se,Chambers:2008gu,Bond:2009xx} it was shown how it can be used, with the separate universe approximation, to compute curvature perturbations.

The $\sigma$ and $\chi$ fields are taken to be position-dependent, and the scale factor $a$ is homogeneous over the whole lattice. The latter assumption is justified when the lattice is smaller than the Hubble volume, and allows us to describe the expansion of the universe using the Friedmann equation.
We have a coupled system of field equations,
\begin{eqnarray}
\label{sigma_eq_motion}
\ddot\sigma+3H\dot\sigma-\frac{1}{a^2}
\nabla^2\sigma+m^2\sigma+g^2\sigma\chi^2&=&0\ ,\\
\label{chi_eq_motion}
\ddot\chi+3H\dot\chi-\frac{1}{a^2}\nabla^2\chi+g^2\sigma^2\chi&=&0\
.
\end{eqnarray}
and the Friedmann equation
\begin{equation}
\label{equ:inhomo3}
H^2=\frac{\overline\rho}{3M_{\rm Pl}^2}\ ,
\end{equation}
where the energy density is calculated as the average energy density in the simulation box,
\begin{equation}
\label{equ:densityint}
\overline\rho=\frac{\rho_{\phi0}}{a^4}+\frac{1}{L^3}\int
d^3x\Bigl[\frac{1}{2}\dot\sigma^2+\frac{1}{2}\dot\chi^2
+\frac{1}{2a^2}\left((\nabla\sigma)^2+(\nabla\chi)^2\right)+V(\sigma,
\chi)\Bigr]\ .
\end{equation}
The coupled system of equations (\ref{sigma_eq_motion}), (\ref{chi_eq_motion}) and (\ref{equ:inhomo3}) are solved on a comoving lattice with periodic boundary conditions. The inflaton component is included as idealised radiation where $\rho_{\phi0}$ is the density in the $\phi$ field at the beginning of the simulation.
We solve the system in conformal time ($d\tau=a^{-1}dt$) using a fourth order Runge-Kutta algorithm. The system of equations is then
\begin{eqnarray}
\label{sigma_eq_motion_conf}
\sigma''+2\frac{a'}{a}\sigma'-\frac{1}{a^2}
\nabla^2\sigma+a^2\left( m^2\sigma+g^2\sigma\chi^2\right)&=&0\ ,\\
\label{chi_eq_motion_conf}
\chi''+2\frac{a'}{a}\chi'-\frac{1}{a^2}\nabla^2\chi+a^2g^2\sigma^2\chi&=&0\
.
\end{eqnarray}
The second order Friedmann equation is
\begin{equation}
\label{equ:inhomo3_conf}
a''=\frac{1}{6}\frac{\overline\rho-3\overline{P}}{M_{\rm Pl}^2}a^3,
\end{equation}
where the averaged density and pressure are
\begin{eqnarray}
\label{equ:densityint_conf}
\fl
\overline\rho=\frac{\rho_{\phi0}}{a^4}+\frac{1}{L^3}\int
d^3x\Bigl[\frac{1}{2a^2}\left(\left(\sigma'\right)^2+\left(\chi'\right)^2\right)
+\frac{1}{2a^2}\left((\nabla\sigma)^2+(\nabla\chi)^2\right)+V(\sigma,
\chi)\Bigr]\ ,\\
\label{equ:pressureint_conf}
\fl
\overline{P}=\frac{\rho_{\phi0}}{3a^4}+\frac{1}{L^3}\int
d^3x\Bigl[\frac{1}{2a^2}\left(\left(\sigma'\right)^2+\left(\chi'\right)^2\right)
-\frac{1}{6a^2}\left((\nabla\sigma)^2+(\nabla\chi)^2\right)-V(\sigma,
\chi)\Bigr]\ .
\end{eqnarray}
\begin{figure}
 \centering
\subfigure[$\sigma$ field]{\includegraphics[width=6.0cm,angle=-90]{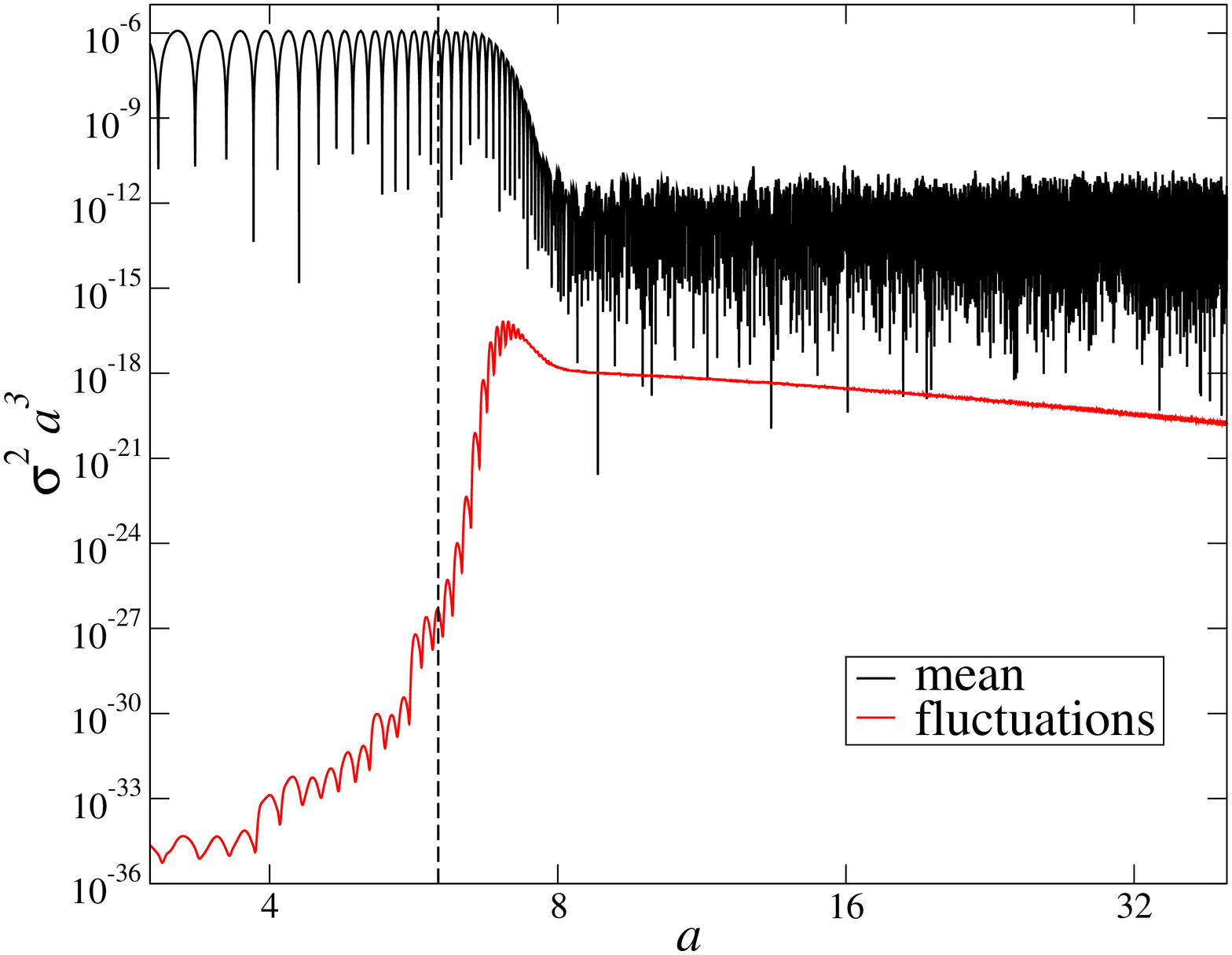}}
\subfigure[$\chi$ field]{\includegraphics[width=6.0cm,angle=-90]{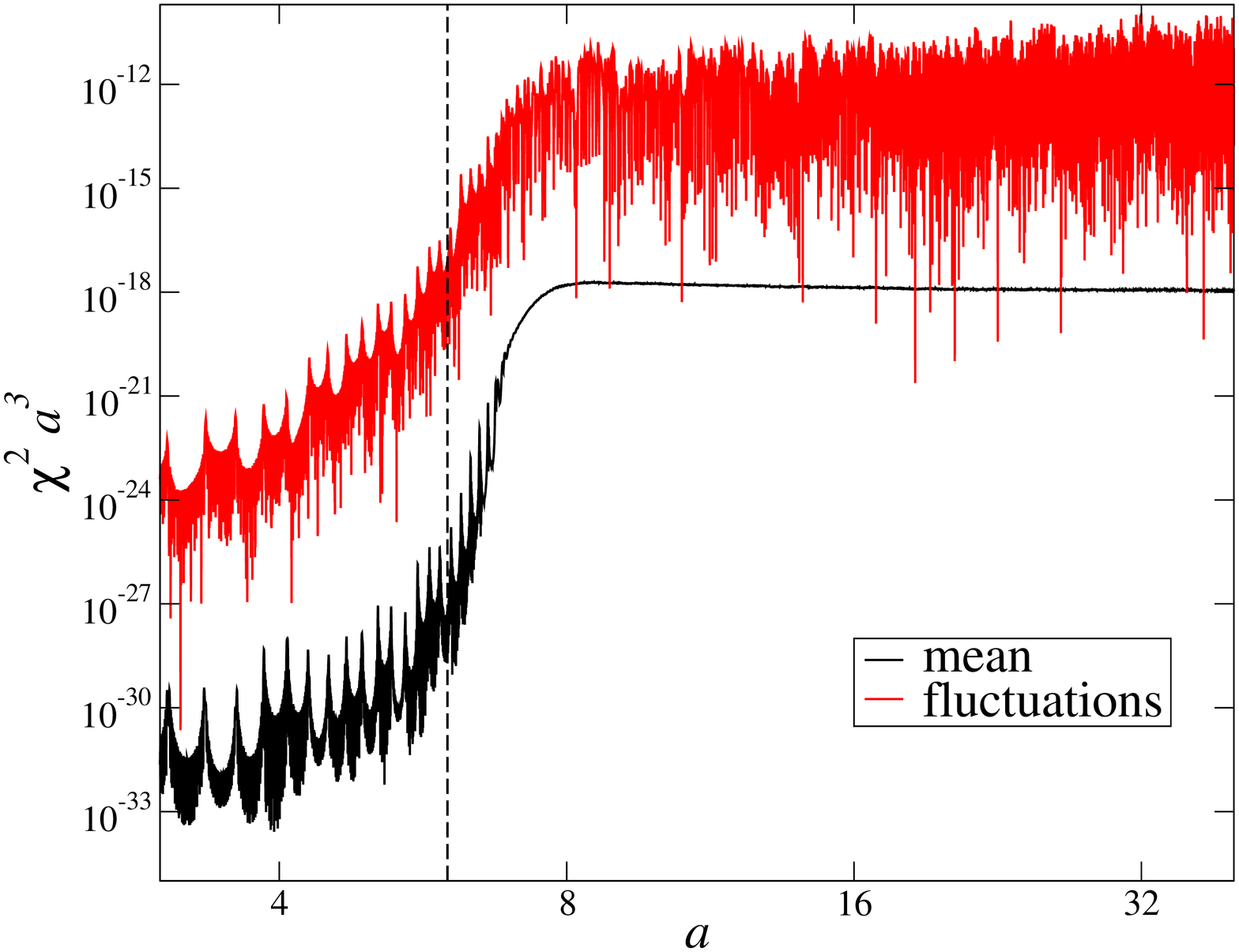}}
 \caption{Evolution of the fields during one simulation run. To the left of the vertical dashed line the evolution is calculated with (\ref{sigma_eq_motion_hom})-(\ref{equ:pressureint_hom}); to the right it is calculated with (\ref{sigma_eq_motion_conf})-(\ref{equ:pressureint_conf}). The $\phi$ field is included as a homogeneous radiation component.}
 \label{fig:a}
\end{figure}

The initial inflaton energy density $\rho_{\phi0}$ depends on the inflaton potential $V(\phi)$.
For simplicity, we assume it has a quartic form,
\begin{equation}
V(\phi)=\frac{1}{4}\lambda\phi^4.
\end{equation}
We choose the initial value of $\phi$ to be $\phi_0=M_{\rm Pl}$,
where we use the subscript $0$ to denote the initial values for the
simulation. We still have the freedom to choose the value of the
conformal time at the start of our simulations. We take
\begin{equation}
\tau_0=\sqrt{\frac{3}{\lambda}}\frac{M_{\rm Pl}}{\phi_0^2}\ ,
\end{equation}
so that the scale factor is simply proportional to conformal time, $a(\tau)\propto \tau$ even at $\tau\sim \tau_0$.

The initial values for the $\sigma$ and $\chi$ fields consist of a homogeneous background component $\sigma_0$ or $\chi_0$, which represents fluctuations with wavelength longer than the lattice size, and
Gaussian inhomogeneous fluctuations which represent short-distance quantum fluctuations.

The $\chi$ field is massive during inflation; its
long-wavelength fluctuations are heavily suppressed.
In contrast, the curvaton field $\sigma$ is still frozen at the start of the
simulation. As a result of superhorizon fluctuations, its local value $\sigma_*$
at the end of inflation varies between one Hubble patch and another; this sets the initial value $\sigma_0$ for the curvaton field in our simulation, $\sigma_0=\sigma_*$.
By running simulations with different initial curvaton values
$\sigma_0$, we can, therefore, determine how $a_{\rm ref}$, $\rho_{\rm
ref}$ and $r_{\rm ref}$ depend on $\sigma_0$ and calculate the curvature
perturbation using (\ref{zeta_exp}).

The relevant range of $\sigma_0$ is determined by the power spectrum $\mathcal P_\sigma(k)$
of the curvaton field.
Assuming that the curvaton mass $m$ is very small, the power spectrum is given by
\begin{equation}
 \label{P_sigma}
 \mathcal P_\sigma(k)\approx\frac{H^2_k}{4\pi^2}\approx\frac{4}{3\pi^2}\lambda M_{\rm Pl}^2N_k^2
\end{equation}
where $H_k$ and $N_k$ are the Hubble rate and the number of
$e$-foldings before the end of inflation evaluated at the time when
the mode $k=a H_k$ left the horizon. A more accurate calculation for non-zero $m$ is given in appendix 1 of \cite{Chambers:2008gu}.

To completely determine the observable curvature perturbation, we need to run the
simulation for all values of $\sigma_*$ that were realised in our current Hubble volume.
The curvaton field is a Gaussian random field, so the mean value over a Hubble volume today,
$\overline{\sigma_*}$, has a Gaussian distribution. The variance of this distribution is
\begin{equation}
\label{equ:varsigmabarN}
\left\langle \overline{\sigma_*}^2\right\rangle =\int_{N_0}^{N_{\rm
tot}}\mathcal P_\sigma\left(k\right)
\left(1-\frac{1}{N_k}\right)dN_k\approx\frac{4}{9\pi^2}\lambda
M_{\rm Pl}^2 N_{\rm tot}^3\ .
\end{equation}
Note that this depends on $N_{\rm tot}$, the total number of
$e$-foldings of inflation, so is essentially a free variable. This
distribution is centred around zero, assuming no
homogeneous classical curvaton background.

The width of the range of $\sigma_*$ that we need to cover is given by the
variance of $\sigma_*$ within our current Hubble volume,
between different Hubble volumes at the end of
inflation. This is given by
\begin{equation}
\label{equ:vardelsigmaN}
\left<\delta\sigma_*^2\right> =\int_0^{N_0}\mathcal
P_\sigma\left(k\right)
\left(1-\frac{1}{N_k}\right)dN_k\approx\frac{4}{9\pi^2}\lambda
M_{\rm Pl}^2 N_0^3\ ,
\end{equation}
where $N_0\approx 60$ is the number of e-foldings after the largest currently
observable scales left the horizon.
This gives the width of the range of $\sigma_*$ that we need to consider. Note
that the only free parameter this depends on is $\lambda$.

For the initial values of the inhomogeneous field modes,
we follow the standard approach~\cite{Khlebnikov:1996mc,Prokopec:1996rr,Felder:2000hq,Rajantie:2000fd,Copeland:2002ku}. The $\chi$ field is given random initial conditions from a Gaussian distribution whose two-point functions are the same as those in the tree-level quantum vacuum state,
\begin{eqnarray}
\label{equ:quantumfluct}
\langle \chi_{\mathbf{k}}^*\chi_{\mathbf{q}}\rangle&=&(2\pi)^3\delta({\mathbf{k}}-{\mathbf{q}} )\frac{1}{2\omega_{\mathbf{k}}}\ ,\nonumber\\
\langle \dot\chi_{\mathbf{k}}^*\dot\chi_{\mathbf{q}}\rangle&=&(2\pi)^3\delta({\mathbf{k}}-{\mathbf{q}} )\frac{\omega_{\mathbf{k}}}{2}\
,
\end{eqnarray}
where $\omega_{\mathbf{k}}=\sqrt{k^2+m_\chi^2}=\sqrt{k^2+g^2\sigma_0^2}.$
All other two-point correlators vanish. The inhomogeneous modes of $\sigma$ are populated similarly to those of $\chi$.

In order to estimate the error bars in, for example, figure 3 we repeat each simulation with the same parameters multiple times using different realisations of the above random initial conditions.
In earlier work on preheating~\cite{Chambers:2007se,Chambers:2008gu,Bond:2009xx},
the whole evolution was calculated using full nonlinear equations. In the current case that would be computationally very expensive, because the model is not conformally invariant and the universe expands by a large factor $\sim 100$ during the evolution. To fit the relevant wavelengths inside the lattice throughout the whole simulation, the lattice size would have to be much larger than $100^3$, which is not possible given the number of simulations which need to be run.

Instead, we take a shortcut, and make use of the fact that until the magnitude of the fluctuations in $\chi$ become large enough to backreact on $\sigma$ (when $\left\langle \chi^2\right\rangle\sim m^2/g^2$) the field dynamics are linear to a very good approximation. Therefore, we can evolve the whole probability distributions of the initial conditions in $k$-space using linearised versions of (\ref{sigma_eq_motion_conf})-(\ref{equ:pressureint_conf}). More precisely,
we need to solve the linear equations for the inhomogeneous modes,
\begin{eqnarray}
\label{sigma_eq_motion_k}
\sigma_{\mathbf{k}}''+2\frac{a'}{a}\sigma'_{\mathbf{k}}
+k^2\sigma_{\mathbf{k}}+a^2 m^2\sigma_{\mathbf{k}}&=&0\ ,\\
\label{chi_eq_motion_k}
\chi_{\mathbf{k}}''+2\frac{a'}{a}\chi'_{\mathbf{k}} +k^2\chi_{\mathbf{k}}+a^2g^2\sigma^2\chi_{\mathbf{k}}&=&0\ ,
\end{eqnarray}
in the background provided by the solutions of the nonlinear equations for the
homogeneous modes,
\begin{eqnarray}
\label{sigma_eq_motion_hom}
\sigma''+2\frac{a'}{a}\sigma'+a^2\left( m^2\sigma+g^2\sigma\chi^2\right)&=&0\ ,\\
\label{chi_eq_motion_hom}
\chi''+2\frac{a'}{a}\chi'+a^2g^2\sigma^2\chi&=&0\ ,
\end{eqnarray}
\begin{equation}
\label{equ:inhomo3_hom}
a''=\frac{1}{6}\frac{\rho-3P}{M_{\rm Pl}^2}a^3\ ,
\end{equation}
\begin{eqnarray}
\label{equ:densityint_hom}
\rho=\frac{\rho_{\phi0}}{a^4}+\frac{1}{2a^2}\left(\left(\sigma'\right)^2+\left(\chi'\right)^2\right)
+V(\sigma,\chi)\ ,\\
\label{equ:pressureint_hom}
P=\frac{\rho_{\phi0}}{3a^4}+\frac{1}{2a^2}\left(\left(\sigma'\right)^2+\left(\chi'\right)^2\right)
-V(\sigma,\chi)\ .
\end{eqnarray}
The background solution is independent of the modes in (\ref{sigma_eq_motion_k}) and (\ref{chi_eq_motion_k}), and therefore it needs to be calculated only once for each $\sigma_0$.

Crucially, the general solution of the linear mode
(\ref{sigma_eq_motion_k}) and (\ref{chi_eq_motion_k}) is obtained by
solving the equations for
two different initial values. This is because, as a consequence of linearity, the general solution can be represented as
\begin{equation}
 \left(\begin{array}{c} \chi_{\mathbf{k}}(\tau)\\ \chi_{\mathbf{k}}'(\tau) \end{array}\right)=M(\tau) \left(\begin{array}{c} \chi_{\mathbf{k}}(0)\\ \chi_{\mathbf{k}}'(0) \end{array}\right)\ ,
\end{equation}
where $M(\tau)$ is a two-by-two matrix. Consider now the solutions with initial conditions of position 1 and velocity 0 and vice versa:
\begin{eqnarray}
 \left(\begin{array}{c} \chi_{\mathbf{k}(1,0)}(\tau)\\ \chi_{\mathbf{k}(1,0)}'(\tau) \end{array}\right)&=&M(\tau) \left(\begin{array}{c} 1\\ 0 \end{array}\right)\ ,\\
 \left(\begin{array}{c} \chi_{\mathbf{k}(0,1)}(\tau)\\ \chi_{\mathbf{k}(0,1)}'(\tau) \end{array}\right)&=&M(\tau) \left(\begin{array}{c} 0\\ 1 \end{array}\right)\ .
\end{eqnarray}
Linearity implies that the solution of arbitrary initial conditions
$\chi_{\mathbf{k}}(0)$ and $\chi'_{\mathbf{k}}(0)$ is given simply by
\begin{equation}
\label{eqn:quickevolve}
 \left(\begin{array}{c} \chi_{\mathbf{k}}(\tau)\\ \chi_{\mathbf{k}}'(\tau) \end{array}\right)=\chi_{\mathbf{k}}(0)\left(\begin{array}{c} \chi_{\mathbf{k}(1,0)}(\tau)\\ \chi_{\mathbf{k}(1,0)}'(\tau) \end{array}\right)+\chi'_{\mathbf{k}}(0)\left(\begin{array}{c} \chi_{\mathbf{k}(0,1)}(\tau)\\ \chi_{\mathbf{k}(0,1)}'(\tau) \end{array}\right)\ .
\end{equation}

We use this approach to evolve the fields until some scale factor $a_1$, which is well before the nonlinear terms become important. Only then, we Fourier transform the fields to coordinate space and start to evolve them using the full nonlinear equations (\ref{sigma_eq_motion_conf})-(\ref{equ:pressureint_conf}). This is shown in figure \ref{fig:a}. We checked that our results do not depend on the chosen value of $a_1$.

This method has two advantages which make the calculation presented here possible. Firstly,
we evolve the whole distribution, so the linear evolution only needs to be calculated once,
and only the relatively short nonlinear stage has
to be repeated for each realisation of the initial conditions.
Secondly, and more significantly, it reduces the factor by which the universe grows during the evolution
of (\ref{sigma_eq_motion_conf})-(\ref{equ:pressureint_conf}) from $\sim$100 to $\sim$10. This allow us to
use lattices with much lower resolution to achieve comparably accurate results, thus significantly
reducing computation time.

\subsection{Simulation results}

\begin{figure}
 \centering
 \includegraphics[width=8cm,angle=-90]{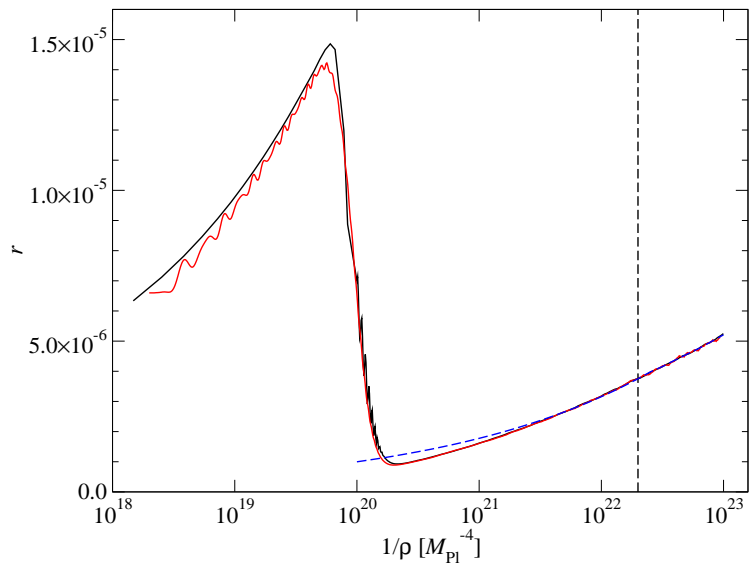}
 \caption{The matter fraction
 $r$ during one simulation. The resonance is rapid and destroys $\sim 95\%$ of the matter. The two solid lines show the two definitions of $r$ given in (\ref{equ:rpressure}) and (\ref{equ:rscalefactor}). $\rho_{\rm ref}=5\times 10^{-23}$ is marked by the vertical dashed line. The blue dashed line represents the approximation (\ref{rho_split}) taken about this $\rho_{\rm ref}$}
 \label{fig:c1a}
\end{figure}

Our model has three free parameters, $\lambda$, $m$, $g$. It is not possible for us to probe this whole
parameter space using current computational technology. Therefore, we simply pick one set of parameters,
$\lambda=10^{-16}$, $g=0.01$ and $m=10^{-8}M_{\rm Pl}$. We take $\chi_0=10^{-16}M_{\rm Pl}$.

We also have to choose the initial value of the curvaton field $\sigma$.
In order to cover all values of $\sigma_*$ in our currently observable universe, we
carried out simulations for a range of $\sigma_0$
\begin{equation}
\overline{\sigma_0}-\frac{1}{2}\delta\sigma_0\le \sigma_0\le
\overline{\sigma_0}+\frac{1}{2}\delta\sigma_0\ ,
\end{equation}
whose width is given by (\ref{equ:vardelsigmaN}),
\begin{equation}
\delta\sigma_0\approx \sqrt{\langle\delta\sigma_*^2\rangle}\approx
10^{-6}M_{\rm Pl}\ .
\end{equation}
The
probability distribution of
the central value $\overline{\sigma_0}$ is given by (\ref{equ:varsigmabarN}),
but as a results of its dependence on the total number of e-foldings $N_{\rm tot}$, which is unknown, we can
essentially choose it freely.
Nevertheless, the value of $\overline{\sigma_0}$ is restricted by several constraints.
The resonance being brought to an end by backreaction demands that $q\gg 1$ (see (\ref{q})).
The characteristic frequency of the $\chi$ field oscillations is $g\sigma_0$ and the time the system
takes
to reach backreaction is proportional to $m^{-1}$. Therefore, the simulation time is proportional to
$g\sigma_0/m=2q^{1/2}$.
In our simulations we used three values:
\begin{equation}
\overline{\sigma_0}=0.0005M_{\rm Pl},\quad 0.001M_{\rm
Pl}\quad\mbox{and}\quad 0.002M_{\rm Pl}\ .
\end{equation}
These corresponds to $q_{br}\approx$ 50, 200 and 750 respectively.
With these parameters one $32^3$ point lattice simulation takes
$\sim$3 hours. We repeat each simulation several ($\sim$50) times with
different random realisations of initial conditions to obtain
statistical errors. Although, this computational method is very
resource intensive, the independence of each simulation (or Hubble
patch) means that the problem scales perfectly on a multiprocessor
machine. A time-step of $(2\times 10^{-4}/m)M^{-1}_{\rm Pl}$ and lattice spacing of $(5m/96)M^{-1}_{\rm Pl}$ were used.

Each simulation gives us time-streams of the scale factor $a$, the
density $\rho$, given by (\ref{equ:densityint_conf}), and
the pressure $P$, given by (\ref{equ:pressureint_conf}). We
measure the matter fraction $r=\rho_m/\rho$ in two different ways:
from the ratio of the pressure and energy density,
\begin{equation}
\label{equ:rpressure}
r=1-3\frac{P}{\rho}\ ,
\end{equation}
and by fitting the dependence of the scale factor on the energy density,
\begin{equation}
\log a=-\frac{1}{4-r}\log\rho+\rm{constant},
\label{equ:rscalefactor}
\end{equation}
in a narrow range about the point of interest.
In continuous time these two expressions would agree, and therefore any discrepancy between them
would be a sign of time discretisation errors.

The measured matter fraction $r$ is shown in figure \ref{fig:c1a} as a function of the energy
density. The two curves correspond to the two definitions (\ref{equ:rpressure}) and
(\ref{equ:rscalefactor}), which clearly agree very well.
Initially, it grows because the matter component decreases
more slowly than the radiation component. At
$\rho^{-1}\approx
10^{20} M_{\rm Pl}^{-4}$,
it drops rapidly as the resonance destroys most of the curvatons. Afterwards $r$ again grows as the
universe expands.

\begin{figure}
\includegraphics[width=12.0cm,angle=-90]{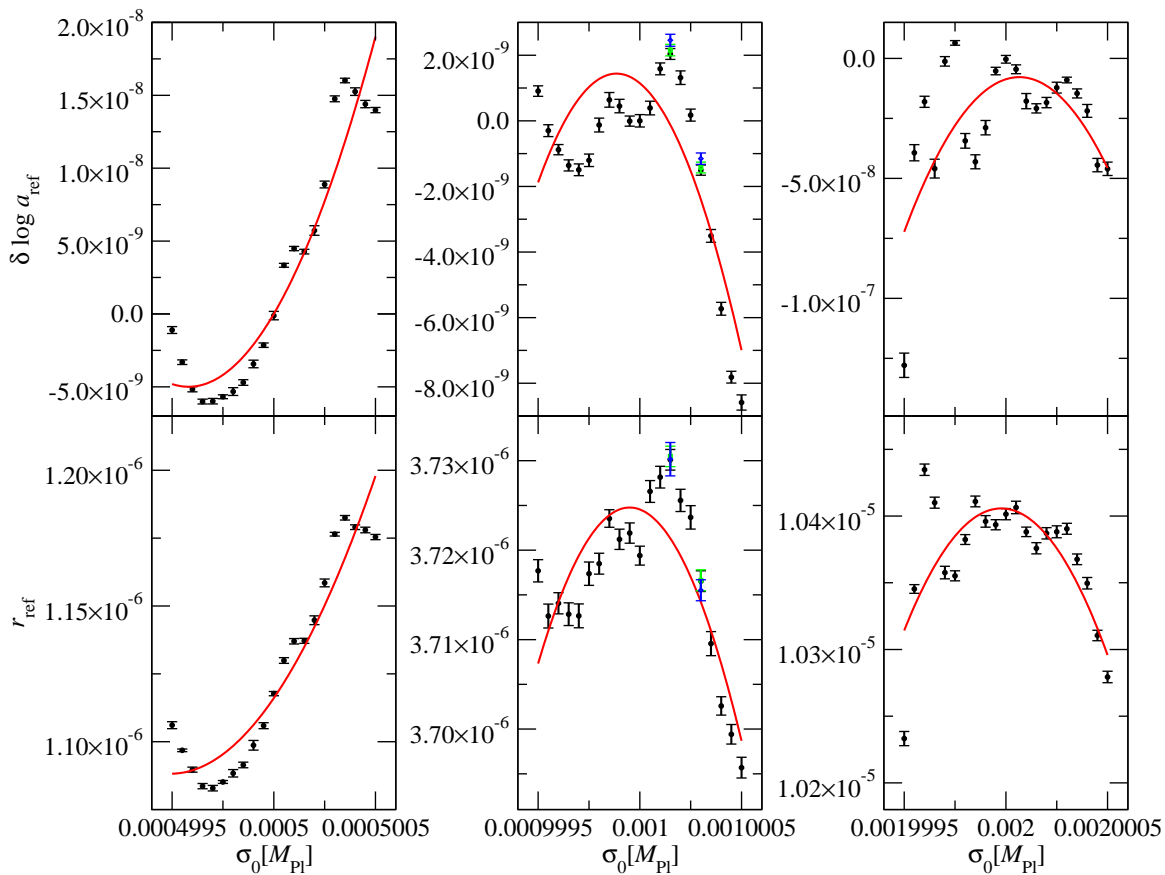}
\caption{Results from simulations for different ranges of $\sigma_0$ centred around $\overline{\sigma_0}=0.0005M_{\rm Pl}$, $\overline{\sigma_0}=0.001M_{\rm Pl}$ and $\overline{\sigma_0}=0.002M_{\rm Pl}$,
each
covering the values present
in one Hubble volume today. The top and bottom rows show $\ln a_{\rm ref}$ and $r_{\rm ref}$ respectively measured
at $\rho=5\times 10^{-23}M_{\rm Pl}^4$ and averaged over $10-50$ runs. The solid lines show quadratic fits of the form
(\ref{equ:fitfunction}). The centre panels also show some numerical checks of the simulation results. The green (square) points represent simulations with time-steps double the length of the black points. The blue (diamond) points are represent a four times longer time-step. It can be seen that the results are within the errors of the primary runs.}
\label{fig:fitplots}
\end{figure}

To calculate the curvature perturbation using (\ref{eqn:lnap}) and (\ref{eqn:lnapp}), we need
to choose a reference point at which we measure the scale factor $a_{\rm ref}$, energy density
$\rho_{\rm ref}$ and matter fraction $r_{\rm ref}$.
We chose this be at fixed energy density $\rho_\refsub =5\times 10^{-23}M_{\rm Pl}^4$.
We interpolated our measurements using a simple least-squares fit in
$\left\{\log(\rho),\ \log(a)\right\}$ and
$\left\{\log(\rho),\ \log(
r)\right\}$ about $\rho=\rho_{\rm ref}$ to find the scale factor $a_{\rm ref}$ and the matter
fraction $r_{\rm ref}$.
The results are shown in figure~\ref{fig:fitplots} for three choices of $\overline{\sigma_0}$.

We found that the choice of $\rho_\refsub $ has no effect on the results as long as it is well after the end of the resonance.
This is demonstrated by figure~\ref{fig:c1a}, in which the (blue) dashed curve shows the matter
fraction $r$ calculated from the assumption~(\ref{rho_split}) of non-interacting matter and radiation components with the measured values of $a_{\rm ref}$ and $r_{\rm ref}$. This agrees
very well with the measured $r(\rho)$ at late times.

We calculate the
first and second derivatives of $\ln a_{\rm ref}$ and $r_{\rm ref}$
with respect to $\sigma_0$.
Assuming that the expansion (\ref{zeta_exp}) works well over the whole range of $\sigma_0$ in the
current Hubble volume, we do this by fitting quadratic polynomials
\begin{eqnarray}
\label{equ:fitfunction}
\ln a_\refsub(\sigma_0)&=&\ln a_\refsub(\overline{\sigma_0}) +\ln a_\refsub'(\sigma_0-\overline{\sigma_0})
+\frac{1}{2}\ln a_\refsub''(\sigma_0-\overline{\sigma_0})^2,\nonumber\\
r_\refsub(\sigma_0)&=&r_\refsub(\overline{\sigma_0}) +r_\refsub'(\sigma_0-\overline{\sigma_0})
+\frac{1}{2}r_\refsub''(\sigma_0-\overline{\sigma_0})^2,
\end{eqnarray}
to our data. These fits are shown by the curves in
figure~\ref{fig:fitplots} and the corresponding fit parameters are given in
table~\ref{tab:fits}.

We concentrate on two measurable quantities: the spectrum of curvature perturbations $\mathcal{P}_\zeta$, which is given by (\ref{eqn:powerzeta}),
and the nonlinearity parameter $f_{\rm NL}$, which is given by (\ref{fnl}).
We chose to fix $\rho_\refsub$ to be a constant, so
(\ref{eqn:lnap}) and (\ref{eqn:lnapp}) simplify to
\begin{eqnarray}
 \label{eqn:lnap_conrho}
 (\ln a)'\Big|_{\rho}&=&(\ln a_\refsub )'+{r_{\rm decay}-r_\refsub  \over 4}{r_\refsub '\over r_\refsub }\ ,\\
 \label{eqn:lnapp_conrho}
 (\ln a)''\Big|_{\rho}&=&(\ln a_\refsub )''+{r_{\rm decay}-r_\refsub  \over 4}{r_\refsub ''\over r_\refsub
 }\ .
\end{eqnarray}
For the amplitude of the curvature perturbations to agree with observations, ${\cal P}_\zeta\approx 10^{-10}$, (\ref{eqn:powerzeta})
and (\ref{eqn:lnap_conrho}) imply
\begin{equation}
\label{eqn:rdec_result}
r_{\rm decay}=r_{\rm ref}+4\frac{r_{\rm ref}}{r'_{\rm ref}}\left[
\pm\sqrt{\frac{{\cal P}_\zeta}{{\cal P}_\sigma}}-(\ln a_{\rm ref})'\right].
\end{equation}
Using this, the nonlinearity parameter (\ref{fnl})
can be recast in the form
 \begin{equation}
 \label{eqn:fnl_result}
 f_{\rm NL} =\frac{5}{6}\frac{\mathcal{P}_\sigma}{\mathcal{P}_\zeta}\left((\ln a_\refsub )''
 +\frac{r_\refsub ''}{r_\refsub '}
 \left(\pm\sqrt{\frac{\mathcal{P}_\zeta}{\mathcal{P}_\sigma}}-(\ln a_\refsub )'\right)\right)\ .
 \end{equation}
The spectrum of the curvaton fluctuations produced during inflation
is given by (\ref{P_sigma}), which, for our parameters is,
$\mathcal{P}_\sigma\approx 5\times 10^{-14}M_{\rm Pl}^2$, so that
$\sqrt{{\cal P}_\zeta/{\cal P}_\sigma}\approx 45$.

The values of $r_{\rm decay}$ and $f_{\rm NL}$ calculated from the fit parameters are shown in
table~\ref{tab:fNL}. In all cases $r_{\rm decay}$ has an acceptable
value, $r_{\rm ref}\ll r_{\rm decay}\ll 1$, which means that the curvaton can produce perturbations
with the observed amplitude. On the other hand,
$f_{\rm NL}$ is much larger than the observations allow~\cite{Komatsu:2008hk,Yadav:2007yy,Komatsu:2009kd,Senatore:2009gt}, which rules out these
parameter values.

\begin{table}
\begin{center}
\begin{tabular}{|c||c|c|c|}

\hline
$\overline{\sigma_0}$  & $0.0005$                          & $0.001$                          & $0.002$ \cr
\hline
\hline
$(\ln a_{\rm ref})'$   & $0.024 \pm 0.002$                 & $-0.005 \pm 0.001$               & $0.025 \pm 0.016$\cr
\hline
$(\ln a_{\rm ref})''$  & $57000 \pm 13000$                  & $-44000 \pm 9000$                & $-420000 \pm 120000$\cr
\hline
$r_{\rm ref}$          & $(1.116 \pm 0.004)\times 10^{-6}$ & $(3.72 \pm 0.002)\times 10^{-6}$ & $(10.4 \pm 0.011)\times 10^{-6}$\cr
\hline
$r_{\rm ref}'$         & $0.109 \pm 0.008$                 & $-0.009 \pm 0.004$               & $-0.014 \pm 0.024$\cr
\hline
$r_{\rm ref}''$        & $(2.1 \pm 0.6)\times 10^5$        & $(1.7 \pm 0.3)\times 10^5$       & $(-8.3 \pm 1.8)\times 10^5$\cr
\hline

\end{tabular}
\end{center}
\caption{Fit parameters defined in (\ref{equ:fitfunction}) in Planck units.}
\label{tab:fits}
\end{table}

\begin{table}
\begin{center}
\begin{tabular}{|c|c|c|c|}

\hline
$\overline{\sigma_0}$      & 0.0005                     & 0.001                       & 0.002 \cr
\hline
\hline
$r_{\rm decay}$            & $0.0018 \pm 0.0001$        & $0.08 \pm 0.03$             & $0.13 \pm 0.21$ \cr
\hline
$f_{\rm NL}$               & $(3.6 \pm 1.0)\times 10^4$ & $(-3.7 \pm 1.7)\times 10^5$ & $(-1.0 \pm 1.8)\times 10^6$ \cr
\hline
\hline
$r_{\rm decay}^{\rm pert}$ & 0.046 & 0.091 & 0.182\cr
\hline
$f_{\rm NL}^{\rm pert}$    & 36 & 18 & 9 \cr
\hline

\end{tabular}
\end{center}
\caption{The values of $r_{\rm decay}$ and $f_{\rm NL}$ calculated using
(\ref{eqn:rdec_result}) and (\ref{eqn:fnl_result}) with the full simulation data (upper) and in the standard
perturbative curvaton theory (lower).}
\label{tab:fNL}
\end{table}

Note that these values of $r_{\rm decay}$ and $f_{\rm NL}$ rely on
the quadratic fit (\ref{equ:fitfunction}), which clearly does not
describe the data very well, as can be seen in figure~\ref{fig:fitplots}.
This illustrates that the curvature perturbations produced in this
model are not well approximated by truncating the expansion
(\ref{fnl_local}) at second order and parameterising the
non-Gaussian effects only by local $f_{\rm NL}$.
Therefore
one should not put too much emphasis on the precise numbers, but the
conclusion that the predicted perturbations are highly non-Gaussian
and ruled out by current observations.

However, our calculations are non-perturbative and
we are not limited to the usual Taylor expansion in (\ref{zeta_exp}).
Instead, using (\ref{eqn:rhosplit_a}) we obtain the whole function $\zeta(\sigma_*)$
as a simple linear combination
\begin{equation}
\label{equ:zetaresult}
\zeta(\sigma_*)={\rm ln}\,a(\sigma_{*})-{\rm
 ln}\,a(\bar{\sigma}_{*})=\delta \ln a_{\rm ref}(\sigma_*)+C\delta r_{\rm
ref}(\sigma_*),
\end{equation}
where the constant $C$ depends on the perturbative curvaton decay rate $\Gamma$ and is
given by
\begin{equation}
C=\frac{1}{4}\left[\left(\frac{\rho_{\rm ref}}{\rho_{\rm decay}}\right)^{1/4}-1\right]
=\frac{1}{4}\left(\frac{r_{\rm decay}}{\overline{r_{\rm ref}}}-1\right).
\end{equation}

\begin{figure}
\begin{center}
\includegraphics[width=10cm,angle=270]{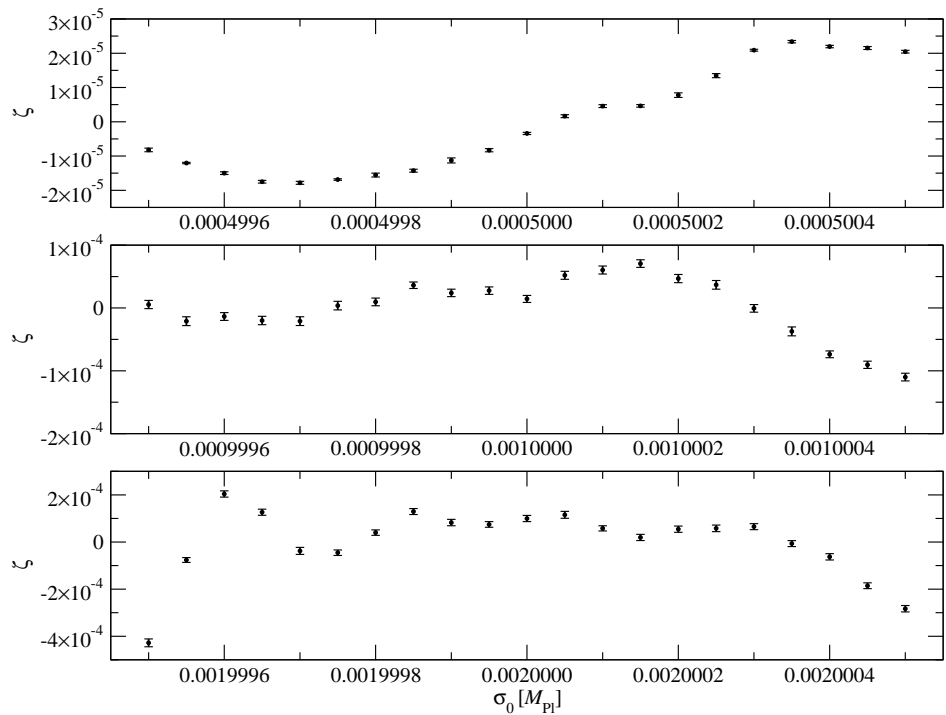}
\end{center}
\caption{
The curvature perturbation (\ref{equ:deltan})
as a function of the curvaton field value for the three cases
(from top to bottom) $\overline{\sigma_0}=0.0005M_{\rm Pl},~0.001M_{\rm Pl},~0.002M_{\rm Pl}$,
calculated using (\ref{equ:zetaresult}) with the value of $r_{\rm decay}$ taken from table~\ref{tab:fNL}. In each case, the dependence is highly nonlinear, implying strong non-Gaussianity.
}
\label{fig:zeta}
\end{figure}

In figure \ref{fig:zeta} we show this full result for the three
choices of $\overline{\sigma_0}$, calculated with the values of
$r_{\rm decay}$ shown in Table~\ref{tab:fNL}. These results give a
complete description of the statistics of curvature perturbations in
this model, and they could be used to produce maps of the CMB or to
compare directly with observations. The correct way to calculate
$f_{\rm NL}$ and other nonlinearity parameters describing higher
order statistics from our results would, therefore, be to repeat the
precise procedure the observers use in their measurements. The data are not well approximated by (\ref{fnl_local})
truncated at second order, so different ways of doing this would
probably yield very different values for $f_{\rm NL}$. Also, $f_{\rm
NL}$ and other nonlinearity parameters would also probably appear
to be scale and position dependent, because on smaller scales one
would cover a smaller range of $\sigma_*$.

Figure \ref{fig:fitplots} shows that, with our parameters, $\delta \ln a_{\rm ref}$ is orders of
magnitude
smaller than the amplitude of the observed perturbations, and, therefore, the dominant contribution to
(\ref{equ:zetaresult}) must come from the second term, i.e. from the variation of the matter
fraction $r_{\rm ref}$. Figure \ref{fig:c1a} shows that the matter fraction drops by a factor of around
$20$ during the nonlinear stage, but in itself, this drop has no effect on the properties of the
final perturbations, because it could be fully compensated for by reducing $\rho_{\rm decay}$.
Instead, what is important is that the amount of curvatons left after the nonlinear stage depends
sensitively on the value of $\sigma_*$. This leads to the variation of $r_{\rm ref}$ within the range of
$\sigma_*$ present in one Hubble volume, which is seen in figure \ref{fig:fitplots} and which makes the dominant contribution to the curvature perturbations.

We believe that the sensitive dependence on $\sigma_{*}$ is caused by the nonlinear field dynamics at the end of the resonance, but we do not have a detailed explanation of this. Nevertheless, it is in qualitative agreement with the results found in \cite{Kofman:1997yn,Podolsky:2005bw} for inflationary preheating, and with
findings that the resonance dynamics in massless preheating are effectively chaotic and that small variations
in initial conditions can have a great impact on curvature perturbations \cite{Bond:2009xx}. In the current case the dependence seems smooth, suggesting that the dynamics are not strictly chaotic.


\section{Conclusions}
\label{sect:conclude}

In this work we have presented a method of calculating the curvature
perturbations from resonant curvaton decay.
The curvaton field is light relative to the Hubble rate during inflation and, as a result,
its fluctuations are correlated on superhorizon scales;
the mean value of the curvaton field varies between one Hubble volume and another.
The local value of the curvaton field affects
both the amount of expansion and the fraction of the curvaton
particles which are destroyed during resonant decay into another field.
This turns fluctuations of the curvaton field into
curvature perturbations, which are generally non-Gaussian. As
the field evolution at the end of the resonance is nonlinear, these
perturbations cannot be calculated using standard perturbative
techniques.

We computed this effect numerically using
classical field theory simulations
combined with the $\delta N$ approach.
This approach was used earlier in refs.~\cite{Chambers:2007se,Chambers:2008gu,Bond:2009xx},
but the current case is technically more difficult because the model is not scale invariant.
To deal with this, we divided the evolution into
three parts: First, the inhomogeneous modes were evolved numerically
using linearised equations. Shortly before nonlinearities become
important, we switched to a full nonlinear simulation to describe
the non-equilibrium dynamics at the end of the resonance.
Eventually, when the system has equilibrated, we
extrapolated to late times by assuming that the universe consists of
non-interacting matter and radiation. This approach allowed us to
study evolution during which the universe expands by several orders
of magnitude, which would not be possible using a straightforward
field theory simulation.

Our method is fully nonperturbative; we are not forced to assume
that the non-Gaussianity is of
the simple form (\ref{fnl_local})
parameterised by $f_{\rm NL}$, and indeed we find that, at least at our parameter
values, it is not.
This may well be a fairly generic consequence of nonlinear field evolution.
Instead, our method produces the full numerical dependence of the curvature perturbation $\zeta$ on
the curvaton field $\sigma_*$, which is a Gaussian random variable with a known power spectrum.
More work is needed to properly compare such data with observations.

Our findings demonstrate that the curvaton resonance can
leave a significant imprint on primordial perturbations. At the parameter values
we studied the perturbations were far too non-Gaussian to be compatible
with observations, but it is very likely that there are other parameter values
with acceptable levels of non-Gaussianity. Therefore it would be interesting to
explore a wider range of parameters in future work.
In this work we focused on parameters for which the curvaton is subdominant throughout the whole
evolution and the decay product field $\chi$ is massive during inflation,
but the other possibilities are equally important and worth
investigating in future work.
We also expect a very similar resonance to take place in a model in which the curvaton is
charged under a gauge group and decays resonantly into gauge bosons.

\section*{Acknowledgements}

This work was supported by the Science and Technology Facilities
Council and by the Academy of Finland grant 130265 and made use of
the Imperial College High Performance Computing
Service\footnote{http://www.imperial.ac.uk/ict/services/teachingandresearchservices/highperformancecomputing}.
The authors would also like to thank Kari Enqvist for valuable
discussions.

\section*{References}
\bibliography{curvaton19a.bib}

\end{document}